\documentclass[preprint,12pt]{elsarticle}

\usepackage{amssymb}
\usepackage{amsmath}
\usepackage{gensymb}
\usepackage{textcomp} 

\journal{Ultramicroscopy}

\begin{document}

\begin{frontmatter}



\title{Single photon emitters in hBN: Limitations of atomic resolution imaging and potential sources of error}


\author[tuwien,universityvienna]{David Lamprecht\corref{cor1}}
\ead{lamprecht@iue.tuwien.ac.at}
\author[universityvienna]{Shrirang Chokappa}
\author[paris]{Alissa M. Freilinger}
\author[universityvienna]{Barbara Maria Mayer}
\author[universityvienna]{Maximilian Melchior}
\author[universityvienna]{Jana Dzíbelová}
\author[universityvienna]{Darwin Lorber}
\author[paris]{Luiz H. G. Tizei}
\author[paris]{Mathieu Kociak}
\author[universityvienna]{Clemens Mangler}
\author[tuwien]{Lado Filipovic}
\author[universityvienna]{Jani Kotakoski}
\cortext[cor1]{Corresponding author}

\affiliation[tuwien]{organization={Institute for Microelectronics, TU Vienna},
            addressline={Gusshausstrasse 27–29}, 
            city={Vienna},
            postcode={1040}, 
            country={Austria}}
\affiliation[universityvienna]{organization={University of Vienna, Faculty of Physics},
            addressline={Boltzmanngasse 5}, 
            city={Vienna},
            postcode={1090}, 
            country={Austria}}
\affiliation[paris]{organization={University Paris-Saclay, CNRS, Laboratoire de Physique des Solides},
            addressline={510 Rue André Rivière}, 
            city={Paris},
            postcode={91405}, 
            country={France}}

\begin{abstract}
There is a growing interest in identifying the origin of single-photon emission in hexagonal boron nitride (hBN), with proposed candidates including boron and nitrogen vacancies as well as carbon substitutional dopants. Because photon emission intensity often increases with sample thickness, hBN flakes used in these studies commonly exceed 30 atomic layers. To identify potential emitters at the atomic scale, annular dark-field scanning transmission electron microscopy (ADF-STEM) is frequently employed. However, due to the intrinsic AA' stacking of hBN with vertically alternating boron and nitrogen atoms, this approach is complicated even in few-layer systems. Here, we demonstrate using STEM image simulations and experiments that, even under idealized conditions, the intensity differences between boron- and nitrogen-dominated columns and carbon substitutions become indistinguishable at thicknesses beyond 17 atomic layers (ca. 6~nm). While vacancy-type defects can remain detectable at somewhat larger thicknesses, also their detection becomes unreliable at thicknesses typically used in photonic studies. We further show that common residual aberrations, particularly threefold astigmatism, can lead to artificial contrast differences between columns, which may result in misidentification of atomic defects. We systematically study the effects of non-radially symmetric aberrations on multilayer hBN and demonstrate that even small residual threefold astigmatism can significantly distort the STEM contrast, leading to misleading interpretations. 
\end{abstract}

\begin{highlights}
\item Single atomic defects are not detectable with ADF-STEM in near-bulk hBN 
\item The appearance of B/N contrast in thick hBN samples is an artifact of aberrations
\item Image processing, contamination and instrumental limitations have to be considered
\item For proper correlation of atomic structures and optical properties advanced methodologies are needed
\end{highlights}


\end{frontmatter}



\section*{Introduction}
Single-photon emitters in hexagonal boron nitride (hBN) have garnered significant attention due to their promise for quantum communication, photonic quantum computing, and quantum sensing applications~\cite{montblanch_layered_2023}. Quantum emitters in hBN commonly exhibit robust, room-temperature emission with high brightness~\cite{gao_high-contrast_2021}, narrow linewidths~\cite{dietrich_solid-state_2020, akbari_lifetime-limited_2022}, and photostability~\cite{li_enhanced_2020}, making it a particularly attractive host material. Despite extensive experimental work on hBN-based quantum emitters~\cite{mendelson_identifying_2021,zhong_carbon-related_2024,chatterjee_room-temperature_2025}, their microscopic origin remains elusive. A wide range of optical emission signatures have been reported, with the most important being the broad emission lines in the near infrared (centered at around 1.5~eV)~\cite{baber_excited_2022}, emission with characteristic phonon sidebands at around 2~eV~\cite{fischer_combining_2023}, a blue photon emission line at around 2.85~eV~\cite{fournier_position-controlled_2021} and ultraviolet (UV) emission at around 4.1~eV~\cite{gale_site-specific_2022}, indicating a heterogeneous population of emitting structures. The ability to unambiguously identify the atomic structures of these emitters is therefore essential, not only for fundamental understanding, but also to enable deterministic defect engineering, a prerequisite for applications in quantum technology. 

Theoretical studies~\cite{cholsuk_hbn_2024}, as well as statistical photon counting~\cite{kumar_localized_2023, patel_room_2024} experiments suggest that the observed emission lines stem from single atomic-scale objects. The most promising candidate for the origin of the near-infrared emissions are charged boron vacancies ($\mathrm{V_B}$)~\cite{grosso_tunable_2017, hayee_revealing_2020}, while carbon substitutional impurities at boron or nitrogen sites ($\mathrm{C_B}$, $\mathrm{C_N}$)~\cite{mendelson_identifying_2021,hayee_revealing_2020}, often in combination with vacancies, interstitial atoms~\cite{zhigulin_stark_2023, ganyecz_first-principles_2024} or other impurity complexes~\cite{kumar_localized_2023, tang_structured-defect_2025}, are considered to be the origin for the higher energy emission lines. However, it is intrinsically challenging to correlate an observed emission with a specific atomic emission site, mostly due to the large difference in resolution between the optical and microscopic techniques (micrometer vs. nanometer scale). An additional complication is, that strong single-photon emission can usually only be detected in hBN flakes approaching the near-bulk limit, with typical thicknesses between 10 and 70~nm~\cite{clua-provost_impact_2024}, corresponding to 30 to 210 layers of hBN. 

In earlier publications, the experimental identification of the emitting sites has mostly been limited to techniques that have a resolution in-between these scales, such as optical near-field techniques~\cite{niehues_nanoscale_2025} or cathodoluminescence measurements~\cite{bourrellier_nanometric_2014}, that allow the localization of the emitter down to a scale of 10~nm. Even though this verifies that the emission is indeed from hBN and not from extrinsic mesoscopic objects decorating the surface, this resolution is not sufficient to clearly identify the origin of the emission, as the observed location can contain multiple different atomic defect sites, that are typically below one~nm in size. Alternatively, some studies employ scanning probe techniques like atomic force and scanning tunneling microscopy. These techniques are indeed able to detect single-atomic defects like C substitutions on the hBN surface~\cite{qiu_atomic_2024}, but their sensitivity is limited to the uppermost atomic layers and thus cannot fully capture the complexity of the emitting sites that may be located deeper within the crystal~\cite{vogl_atomic_2019}. 

To overcome this issue, several recent publications~\cite{singla_direct_2025, singla_probing_2024, hua_deterministic_2025, liang_site-selective_2025, hou_nanometer_2025} study candidates for the emission sites using annular dark-field scanning transmission electron microscopy (ADF-STEM), where the intensity of each atomic column scales approximately with the square of the atomic number $Z$~\cite{krivanek_atom-by-atom_2010}, making it a potentially powerful tool for detecting vacancies (which have reduced intensity) and substitutional carbon dopants (which lead to a decrease or increase in intensity depending on the site they substitute), irrespective of their location in the depth direction. 

However, interpreting ADF-STEM images of multilayer hBN is challenging due to its AA' stacking order, which causes alternating contrast patterns that depend on the parity of the layer number. As thickness increases, differentiating boron- and nitrogen-dominated columns—as well as subtle defects like carbon substitutions—becomes increasingly difficult due to diminishing contrast difference and increasing noise levels. Moreover, instrumental and imaging limitations such as probe aberrations further complicate defect identification, especially when such effects can introduce periodic contrast changes that may not be detected by pure visual inspection~\cite{sawada_accurate_2017, lopatin_aberration-corrected_2020}, raising concerns about the reliability of using ADF-STEM for defect identification in multilayer hBN. 

In this work, we systematically investigate the limitations of ADF-STEM imaging for structural analysis of single-photon emitter candidates in multilayer hBN using detailed image simulations, focusing on the role of hBN thickness and residual aberrations, and show that even under idealized favorable conditions, single substitutional carbon atoms can become invisible at hBN thicknesses of about 3~nm. We demonstrate both through simulations and experiments that even small residual threefold astigmatism can significantly distort the STEM contrast, leading to misleading interpretations such as artificial B/N contrast in ADF-STEM images of near-bulk hBN. Finally we also provide practical example of thickness-dependent cathodoluminescence (CL), and show that for proper correlative studies, methodical refinements are required. 

\section*{Results and Discussion}
\subsection*{Detecting point defects in multilayer hBN}

Several recent studies have attempted to identify the presence of carbon atoms in hBN by analyzing histograms of atomic column intensities in ADF-STEM images~\cite{singla_probing_2024, singla_direct_2025, hua_deterministic_2025, hou_nanometer_2025}. In these analyses, atomic columns with intensities between the main peaks corresponding to boron and nitrogen have been interpreted as containing carbon. This method is valid for monolayer hBN and has been demonstrated in several publications~\cite{sohlberg_insights_2015,park_atomically_2021}. However, exfoliated multilayer hBN has a natural AA' stacking order, that leads to periodic modulation in projected column intensities depending on the parity of the layer number, with even layer numbers having no difference between the columns with a boron ("B") and nitrogen ("N") in the bottom layer (seen from the direction of the beam); and odd layer numbers exhibiting the known alternating pattern. However, when the number of layers increases, the identification of boron- and nitrogen dominated sites in odd-layered hBN becomes increasingly difficult, as the noise level arising from various instrumental and physical limitations quickly exceeds the increasingly subtle intensity difference between the columns. 

\begin{figure}[htp]
    \centering
    \includegraphics[width= 14cm]{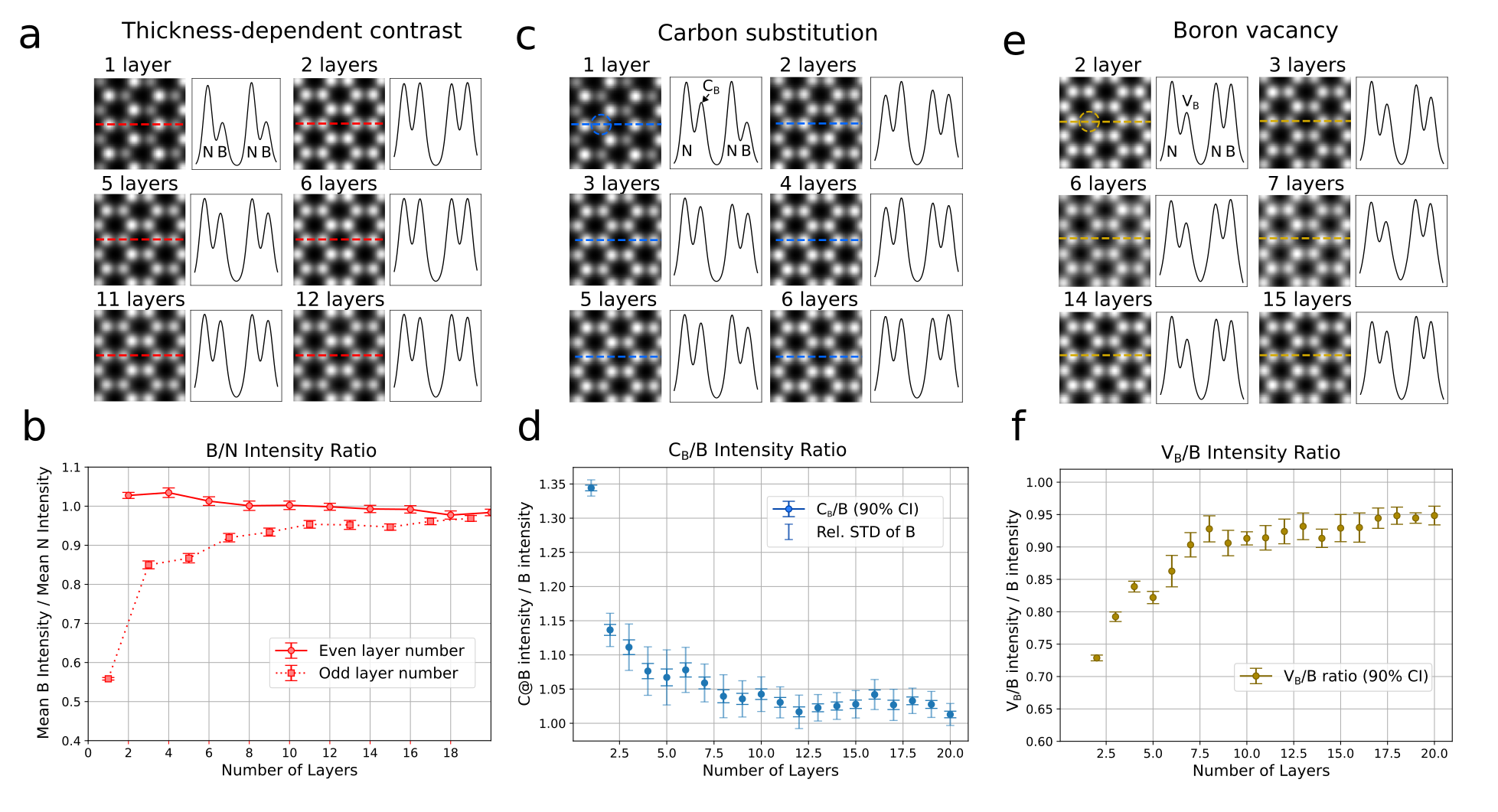}
    \caption{Thickness-dependent visibility of atomic features in HAADF-STEM images in hBN. \textbf{a)} Simulated images of hBN with different layer numbers and line profiles. The labels N and B refer to the atoms in the bottom layer in the respective atomic column. \textbf{b)} Average  intensity ratio of columns with a boron ("B") and nitrogen ("N") in the bottom layer as a function of layer number, for even layer numbers (solid line) and odd layer numbers (dotted line); the error bars (mostly within the markers) represent the 90 \% confidence interval of the ratio. \textbf{c)} Simulated images of hBN with different layer numbers with a carbon atom replacing one boron atom, as well as line profiles over substituted and pristine atomic column pairs. \textbf{d)} Intensity ratio between a pristine "B" column and a carbon implanted "B" column. The dark blue error bars represent the 90 \% confidence interval of the ratio, the light blue error bars represent the standard deviation of the "B" columns relative to the ratio.  \textbf{e)} Simulated images of hBN with different layer numbers with a boron vacancy in the "B" column as well as line profiles over defective and pristine atomic column pairs. \textbf{f)} Intensity ratio between a defective and pristine "B" column as a function of layer number. The error bars represent the 90 \% confidence interval of the ratio.
}
    \label{Fig 1}
\end{figure}

To demonstrate this, we performed multislice STEM image simulations of hBN with increasing thicknesses using the open-source simulation tool \textit{ab}TEM~\cite{madsen_abtem_2021}. Simulations offer the advantage of isolating and controlling specific variables without the influence of instrumental imperfections or unintentional experimental bias, which we will later show to be a potentially critical factor. Fig.~\ref{Fig 1}a shows simulated ADF-STEM images of pristine hBN ranging from 1 to 20 layers (ca. 0.3 to 7~nm thickness). These simulations were performed under idealized conditions: a kinetic energy of 60 keV, no aberrations, a probe size of 0.03~nm, and thermal diffuse scattering, approximated by the frozen phonon model (see Methods). Even from a simple visual inspection of odd-layer examples, it is apparent that the intensity difference between atomic columns decreases rapidly with increasing thickness.

This observation is quantified in Fig. \ref{Fig 1}b, which shows the intensity ratio between "B" and "N" columns as a function of layer number, separately for even- and odd-layered hBN. Odd-layered samples (e.g., 1, 3, 5 layers) retain a measurable intensity difference between "B" and "N" columns. However, beyond five layers, the noise introduced through thermal diffuse scattering begins to dominate, and by 17 layers the "B"/"N" intensity ratios of odd- and even-layered samples overlap, making them effectively indistinguishable, even under unrealistically ideal imaging conditions. 

Equivalent simulations and $\mathrm{C_B}$/"B" intensity ratio plots for hBN with carbon substitutions in the bottom layer are presented in Fig.~\ref{Fig 1}c–d, where single carbon atoms randomly replace boron atoms in the bottom layer, representing a defect configuration commonly suspected to cause blue photon emission in hBN. The simulated images and corresponding $\mathrm{C_B}$/"B" intensity ratios (Fig.~\ref{Fig 1}d) show that even under these ideal conditions, identifying individual carbon atoms becomes highly unreliable already beyond about 7 layers. This finding challenges recent reports of single-carbon atom identification in hBN flakes as thick as 210 layers~\cite{singla_probing_2024}.

In Figs.~\ref{Fig 1}e-f, we also show the effect of increasing layer number on the visibility of single boron vacancies ($\mathrm{V_B}$). While, under simulated conditions, $\mathrm{V_B}$ can be identified up to the maximum displayed thickness of 20 layers, the trend clearly shows that their detection becomes unreliable at thicknesses above 10~nm (30 layers) — comparable to the thicknesses commonly used in photonic quantum emitter studies. Note that this applies only to individual isolated vacancies, as defect clusters lead to lower intensity areas, which can be distinguished more easily via visual inspection.

\begin{figure}[htp]
    \centering
    \includegraphics[width= 14cm]{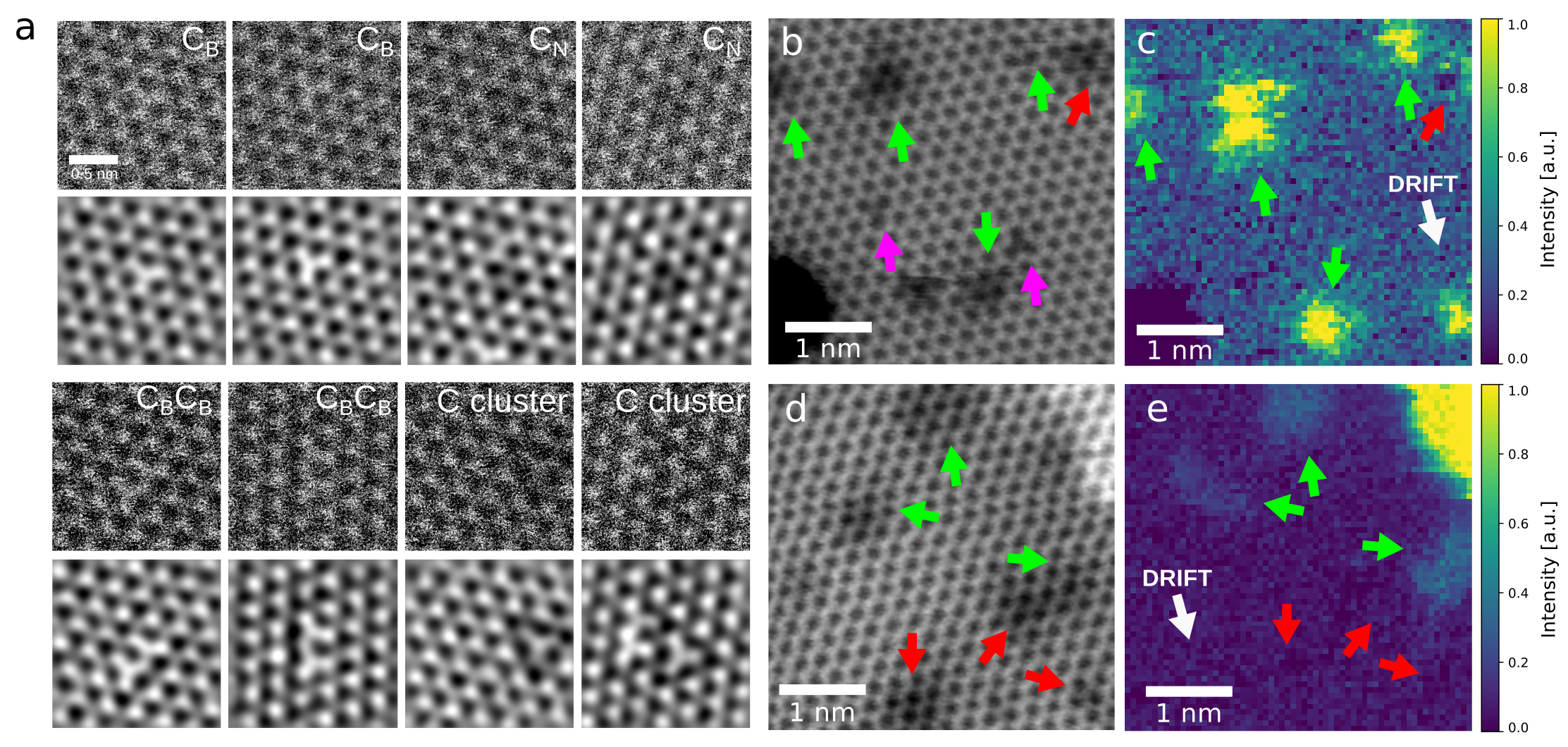}
    \caption{Detection of carbon substitutions using HAADF-STEM and EELS. \textbf{a)} Raw and double Gaussian filtered images of different carbon features found in monolayer hBN. \textbf{b)} Experimental MAADF-STEM image of bilayer hBN with clusters of carbon substitutions. Green arrows mark clusters with corresponding EELS carbon K-edge intensity peaks, purple and red arrows mark single defects and clusters without such an intensity peak. \textbf{c)} Corresponding EELS map integrated around the carbon K-edge energy loss region (290-310~eV energy loss). To increase the visibility of the carbon edge the highest and lowest 5 percentile of integrated intensity values have been clipped. The white arrow indicates the approximate direction of drift during EELS acquisition. \textbf{d)} Equivalent experimental MAADF-STEM image of a six-layer hBN sample with clusters of substitutional carbon. Green arrows mark clusters with corresponding EELS carbon K-edge intensity peaks, red arrows mark clusters without such an intensity peak. \textbf{e)} Corresponding EELS map integrated around the carbon K-edge energy loss region (290-310~eV energy loss). To increase the visibility of the carbon edge the highest and lowest 5 percentile of integrated intensity values have been clipped. The white arrow indicates the approximate direction of drift during EELS acquisition.
}
    \label{Fig 2}
\end{figure}

To demonstrate the detectability of carbon atoms implanted into the hBN lattice, we imaged monolayer chemical vapor deposition (CVD) grown and multilayer exfoliated hBN using ADF-STEM and electron-energy-loss-spectroscopy (EELS) mapping. Fig.~\ref{Fig 2}a shows raw medium-angle ADF-STEM images (60-200 mrad, MAADF) as well as double Gaussian-filtered images of substitutional carbon atoms in monolayer hBN. The experimental parameters of this acquisition were chosen to replicate conditions of typical quantitative defect studies, where the total dose has to be kept low to avoid the introduction of additional defects during imaging~\cite{bui_creation_2023}. While single carbon substitutions can be reliably detected both in the raw and filtered images, the defect structure of larger clusters remains ambiguous even in monolayer hBN. Fig.~\ref{Fig 2}b,c show MAADF images and corresponding EELS maps integrated around the carbon K-edge (290-310~eV energy loss) of an exfoliated bilayer hBN sample where carbon atoms have been implanted by leaking methane into the microscope column and using the electron beam to introduce vacancies that subsequently get filled by carbon atoms originating from methane dissociation~\cite{wei_electron-beam-induced_2011, park_atomically_2021, mayer_electron-beam-induced_2025}. In MAADF images certain lower-contrast regions (marked by green arrows) can be clearly discerned from the pristine lattice and a comparison with the carbon K-edge intensity convincingly proves that these regions contain carbon atoms. Note that the positions of the clusters are slightly shifted from the EELS map to the MAADF image due to slow drift during the EELS acquisition. Two single-atom defects (purple arrow) and a larger defect cluster (red arrow), clearly visible in the MAADF image have no corresponding peak in the EELS intensity map, indicating that these consist out of $\mathrm{V_B}$ or vacancy aggregates. The comparison shows that carbon atoms in bilayer hBN can be distinguished from the pristine lattice, even though additional spectroscopic evidence may be necessary to reliably differentiate between carbon substitutions and vacancies. Figs.~\ref{Fig 2}d-e show similar images collected on a six layer hBN sample. Here, some clusters with lower intensity (green arrows) in the MAADF image have corresponding maxima in the carbon K-edge EELS map, while other lower intensity features (red arrows) do not, despite exhibiting similar reductions of intensity. These results suggest that with increasing thickness, confidently identifying whether a feature arises from implanted carbon, vacancies, or from a combination of both, becomes increasingly challenging. We note however, that at least in monolayer hBN, single carbon atoms can be reliably identified using a more specialized setup for correlated atomic-scale HAADF and EELS mapping, as we showcase in Supplementary Materials Fig.~\ref{Luiz EELS}.

\subsection*{Artificial B/N contrast in multilayer hBN}

We now address the second critical question: Why do some microscopy images show an apparent clear intensity difference between alternating atomic columns in up to 100~nm thick hBN samples? 
While ADF-STEM images can be affected by a range of instrumental imperfections, most of these introduce either uniform or patterned, noise-like effects, such as mechanical vibrations, drift, and electron shot noise, or they produce radially symmetric effects (for example defocus, spherical (Cs) or chromatic aberrations), which primarily "blur out" the contrast and reduce overall resolution. In contrast, non-radially symmetric aberrations such as astigmatism (A12), coma (A21), and threefold astigmatism (A23) can modulate the intensity distribution of individual atomic columns without causing a necessarily obvious degradation in resolution. This makes them significantly more difficult to identify and correct in standard imaging workflows. These aberrations introduce angle dependent phase shifts in the probe wavefunction, which is why they are best described in polar coordinates, where each aberration is characterized by a magnitude $C_{xx}$ and a phase $\phi_{xx}$. 

\begin{figure}[htp]
    \centering
    \includegraphics[width= 14cm]{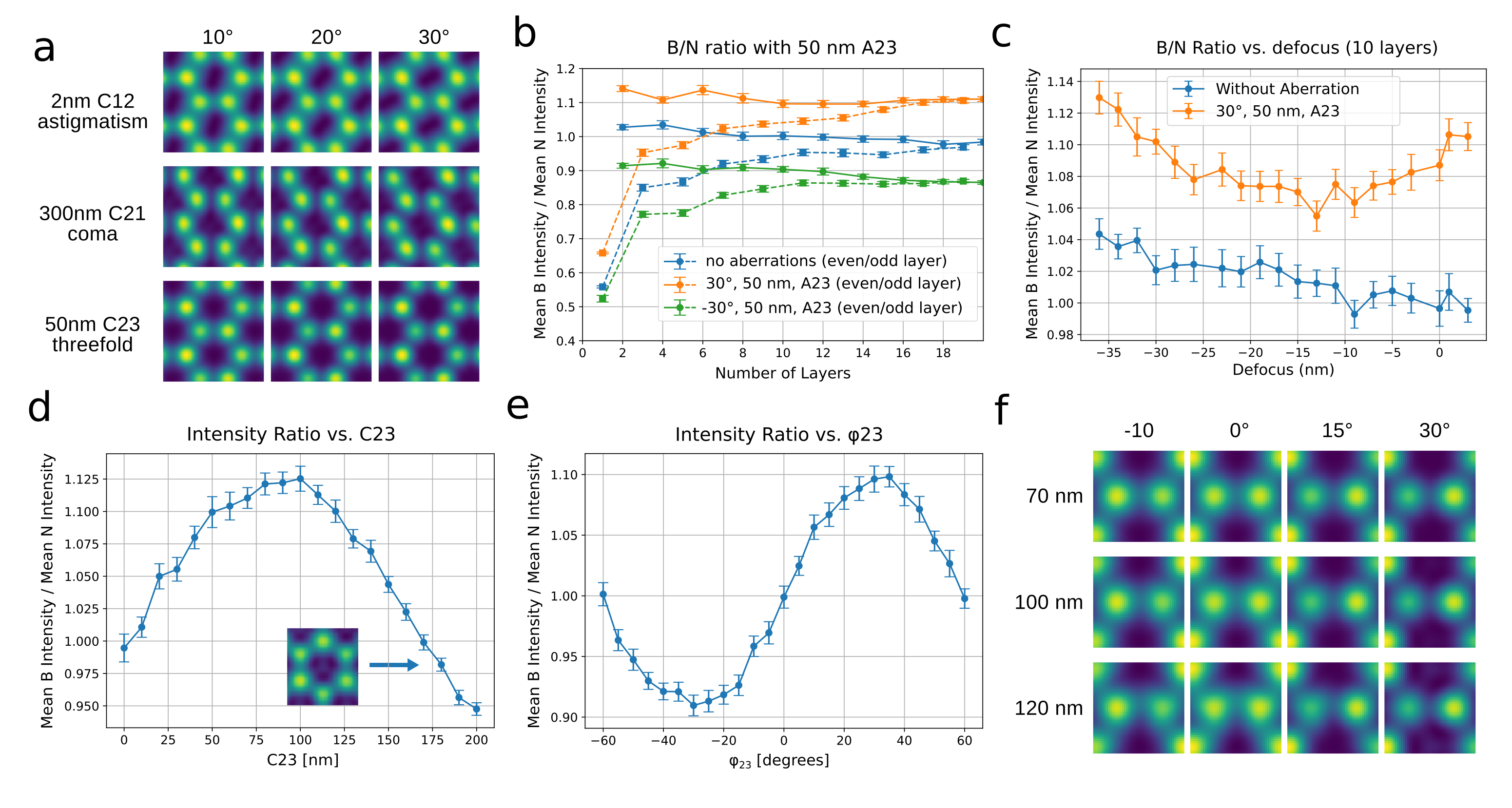}
    \caption{Effects of non-centrosymmetrical aberrations on B/N intensity ratios. \textbf{a)} Simulated images of bilayer hBN showcasing the typical effect of astigmatism, coma and threefold astigmatism for different relative angles between crystal direction and aberration phase. \textbf{b)} Intensity ratios of neighboring columns as a function of layer number for aberration-free settings, with 50~nm A23 aberration in phase with the crystal direction and with 50~nm A23 in counter-phase with the crystal direction. \textbf{c)} Intensity ratios of 10-layer hBN for aberration-free and 50~nm in-phase A23 aberration settings as a function of defocus. \textbf{d)} Intensity ratio as a function of A23 amplitude ($C_{23}$) for the in-phase condition. Inset: Simulation showing the artificial "AB-stacking" effect at higher amplitudes. \textbf{e)} Intensity ratio as a function of A23 phase ($\phi_{23}$) at a fixed A23 amplitude of 50~nm. \textbf{f)} Simulated images showcasing the effect of strong threefold astigmatism at different amplitudes and phases.} 
    \label{Fig 3}
\end{figure}

To explore the specific effects of these aberrations, we simulated ADF-STEM images of bilayer hBN containing neighboring boron and nitrogen atoms under various non-radially symmetric aberration conditions. Fig.~\ref{Fig 3}a illustrates the influence of A12 ($C_{12}$ = 2~nm), A21 ($C_{21}$ = 300~nm), and A23 ($C_{23}$ = 50~nm), for out-of-phase (10° and 20°) and in-phase (30°) offsets with the crystal orientation. As shown, all aberrations, but especially A23, induce variations in the intensity ratio of neighboring atomic sites. However, A12 and A21 also distort the apparent symmetry of atomic columns, producing visually detectable artifacts. This makes them easier to recognize and compensate for during alignment.

Threefold astigmatism (A23), particularly when its phase aligns with the crystal orientation, poses a more subtle problem. It introduces pronounced contrast modulation between the atomic columns without visibly distorting the apparent shape or symmetry of the atomic columns themselves. This makes its presence easy to overlook, even in high-resolution images. The origin of the effect lies in the symmetric coupling between the threefold aberration and the three- or sixfold symmetric potential of the hexagonal lattice, a phenomenon that is well established in electron microscopy and has already been studied in the context of monolayer graphene~\cite{lin_effects_2015, hofer_atom-by-atom_2021} and transition metal dichalcogenides~\cite{lopatin_aberration-corrected_2020}. 

To test the hypothesis that even a small amount of A23 can lead to significant artificial contrast in neighboring atomic columns, we now move to a systematic simulation-based analysis of the influence of threefold astigmatism on ADF-STEM contrast in multilayer hBN. We  first study the effect of A23 as a function of layer number. We use a $C_{23}$ amplitude of 50~nm, which is well within the range of typical residual aberrations in well-aligned aberration-corrected STEM systems. The phase angle $\phi_{23}$ is set either in-phase (30\degree~offset) or counter-phase (-30\degree~offset) with respect to the crystal orientation. The evolution of the intensity ratio as a function of thickness for pristine hBN, compared to cases with ±30\degree~A23 aberration, is shown in Fig.~\ref{Fig 3}b. When the aberration is applied in-phase (30\degree), we observe a systematic increase in the 
intensity ratio across all layer numbers. This shift not only reverses the apparent intensity contrast for even-layered samples but also flattens the ratio in odd-layered hBN. For example, at 50~nm amplitude, 7- and 9-layered hBN, which should exhibit distinguishable intensity differences, instead have intensity ratios near unity and can therefore be easily mistaken for even-layered samples. In contrast, a counter-phase offset (-30\degree) leads to the opposite effect: over-enhancing the N-dominated columns and making all thicknesses resemble monolayer hBN with artificially enhanced contrast. 

As can be seen in Fig.~\ref{Fig 1}b in the aberration-free case, the "B"/"N" ratio is not exactly even for even layer numbers. This is due to the combination of two effects: First, as the electron beam spreads over a larger area after every scattering event, layers closer to the entry-side contribute more strongly to high-angle scattering. This leads to a slight increase in the intensity ratio for lower layer numbers. For thicker samples, this effect becomes overshadowed by the stronger overall scattering. Here, however, the intensity ratio changes depending on the defocus value. Fig.~\ref{Fig 3}c demonstrates this for a 10-layer system: For both the aberration-free and the 50~nm in-phase threefold astigmatism case, we see a significant and not necessarily linear dependence on defocus. Since no universal “optimal” defocus exists for thick samples, we therefore fixed the defocus to the bottom layer (zero defocus) in all simulations.

In Fig.~\ref{Fig 3}d, we further analyze the intensity ratio as a function of $C_{23}$ amplitude at a fixed phase angle of 30\degree. The contrast shift increases steadily with aberration strength, peaking at around 120~nm. Beyond this point, the intensity ratio begins to decline again. This non-monotonic behavior arises from the increasingly elongated tails of the electron probe at higher aberration amplitudes. At intermediate amplitudes, A23 shifts intensity from one atomic column to its neighbor, enhancing contrast. At higher amplitudes, however, a significant portion of the intensity is transferred into the atom-free region in the middle of the hexagonal rings, reducing contrast between adjacent columns and leading to contrast inversion at very high amplitudes. At around 180~nm amplitude, this redistribution of intensity results in an image that resembles AB-stacked hBN, as illustrated in the inset of Fig ~\ref{Fig 3}d.

Finally, we study the dependence of the intensity ratio on the phase angle $\phi_{23}$ at a fixed amplitude of 50~nm. Fig.~\ref{Fig 3}e reveals a sinusoidal relationship between the phase angle and the intensity ratio. Nearly all phase offsets, except for narrow angular windows near multiples of $\phi_{23}$ = 60\degree, lead to significant contrast shifts. Fig.~\ref{Fig 3}f shows simulated ADF-STEM images of 10-layer hBN for a range of high $C_{23}$ and $\phi_{23}$ values. These visualizations confirm that even at high amplitudes, off-resonant A23 introduces only minor distortions of the atomic columns, in contrast to the more pronounced symmetry breaking caused by A12 and A21. Nonetheless, even small near-resonant A23 components can strongly alter the intensity ratio, making it an especially insidious source of error in defect studies. 

\begin{figure}[htp]
    \centering
    \includegraphics[width= 14cm]{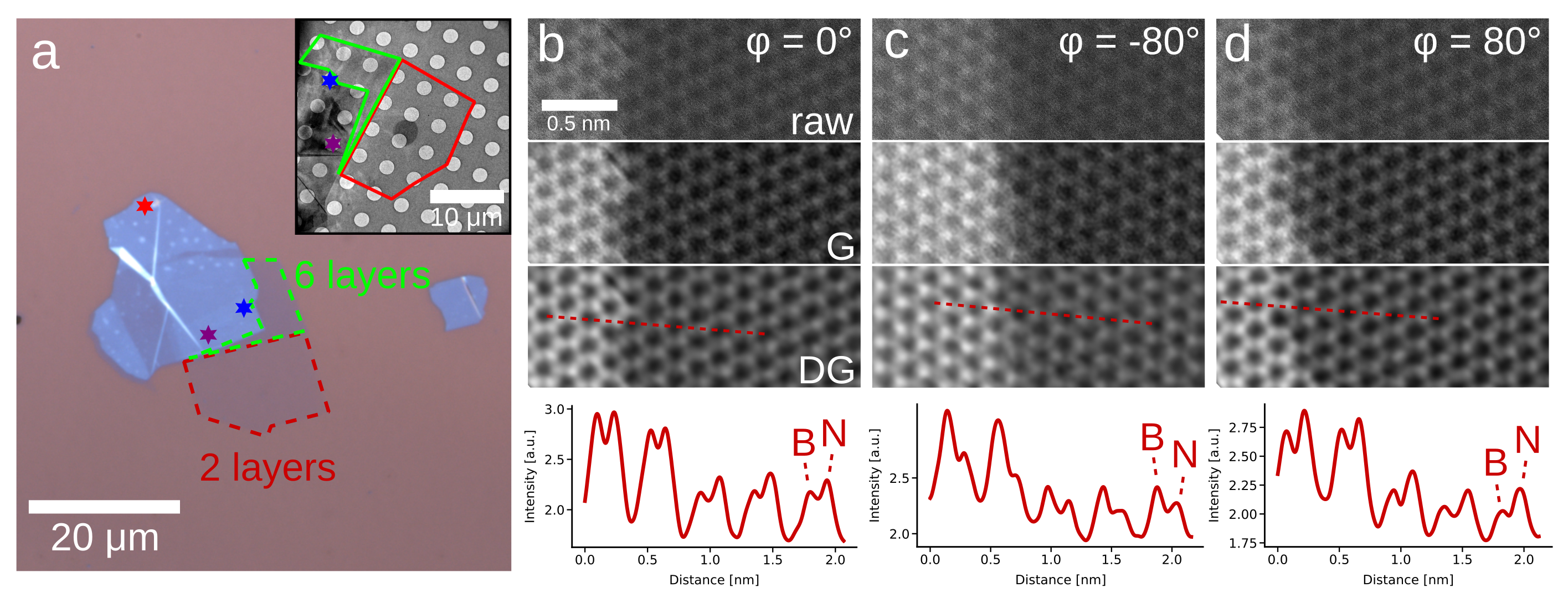}
    \caption{Experimental validation for A23-related effects. \textbf{a)} Light microscope image of the exfoliated hBN flake on a 280~nm SiO$_2$/Si substrate. The red, purple and blue stars denote the locations in the near-bulk where the images in Fig. \ref{Fig 5}d-g have been obtained. Inset: Overview bright field electron microscopy image of the flake after transfer to the TEM grid obtained with the CCD camera. \textbf{b)} Raw, Gaussian-filtered (G) ($\sigma$ = 3 px) and double Gaussian-filtered (DG) (bandpass around first order FFT peaks, inner/outer radius 3.0/6.0~nm$^{-1}$) ADF-STEM images of a bilayer-monolayer step region without additional A23 aberration as well as a line profile over the bilayer-monolayer step in the double Gaussian-filtered image. The scale bar in the first image is valid for all images. \textbf{c)}  Equivalent images from the same region, but with additional ($C_{23}$ = 132~nm, $\phi_{23}$ = -80\degree) corrector detuning. \textbf{d)}  Equivalent images from the same region but with ($C_{23}$ = 132~nm, $\phi_{23}$ = +80\degree) corrector detuning.} 
    \label{Fig 4}
\end{figure}

\subsection*{Experimental replication of artificial B/N contrast}

To complement our A23 simulation results, we also performed ADF-STEM imaging of a mechanically exfoliated hBN flake containing regions of varying thickness. The sample shown in Fig. \ref{Fig 4}a was first exfoliated onto a 300~nm SiO$_2$/Si substrate, then transferred onto a Quantifoil TEM grid with an amorphous carbon support film (see inset of Fig.~\ref{Fig 4}a). Thickness determination for the two thinner regions was carried out by drilling nanoscale holes using the 60 keV electron beam of the Nion UltraSTEM 100 (see Supplemental Materials Fig.~\ref{drilling}). Based on this method, the region marked by a red border was identified as bilayer hBN, while the region marked in green corresponds to six layer hBN.

For the thicker, near-bulk regions (marked by stars in Fig.~\ref{Fig 4}a), we estimated the number of layers by comparing experimental position averaged convergent beam electron diffraction (PACBED) patterns with simulated patterns~\cite{lebeau_position_2010} (see Methods). This comparison yielded optimal results for 16-18 layers. Additionally, we collected low-loss EEL spectra at all marked positions, which allows for robust thickness estimation using the $t/\lambda$ method~\cite{iakoubovskii_thickness_2008} calibrated against the known 2- and 6-layer regions (see Methods). This method also yielded approximate thicknesses of ca. 16 layers. The results of both thickness estimation methods are shown in Supplementary Figs.~\ref{PACBED} and~\ref{EELSthickness}. 

Fig.~\ref{Fig 4}b presents raw, Gaussian-filtered and double Gaussian-filtered ADF-STEM images of the contamination-free bilayer region, where the upper layer has been partially removed by electron irradiation, acquired after automatic aberration correction up to fifth order, followed by manual adjustment of defocus, A12, and A21. The qualitative accuracy of the correction is supported by the presence of clear "B"/"N" contrast in the monolayer region, and the absence of such contrast in the bilayer, consistent with expectations from our simulations. 

To demonstrate the effect of threefold astigmatism A23, we introduced controlled aberrations by manually adjusting the Cartesian A23a and A23b components in the corrector-tuning interface. Fig.~\ref{Fig 4}c shows the same region imaged with a detuning corresponding to an A23 amplitude of $C_{23}$=~132~nm and phase $\phi_{23}$=~-80\degree. Under this setting, the column contrasts invert, with the region to the left of the boron–nitrogen atomic column pair ("B"-site) exhibiting significantly higher intensity in both mono- and bilayer. When Fig.~\ref{Fig 4}d the A23 phase is reversed (Fig.~\ref{Fig 4}d, $R_{23}$=~132~nm, $\phi_{23}$=~80\degree), the effect switches direction, with enhanced intensity now on the right side of the atomic column pair ("N"-site).

Interestingly, the observed intensity changes are asymmetric: the enhancement on the "B"-site is significantly more pronounced, and the image has a slightly lower resolution compared to the neutral state, while the "N"-site exhibits less enhancement but appears sharper. This asymmetry can be attributed to several factors. First, the baseline setting (Fig.~\ref{Fig 4}a) will still contain residual, uncorrected non-centrosymmetric aberrations that slightly favor one atomic site, which cannot be trivially excluded. Second, our phase alignment is only a rough approximation of the actual resonance condition; achieving true alignment would require a sample holder with in-plane rotational freedom, as previously demonstrated by Lopatin et al. \cite{lopatin_aberration-corrected_2020}. Simply circulating $\phi_{23}$ with corrector detuning is not a fully equivalent alternative, because the correction function is multi-dimensional and highly non-linear, which means that changing the setting of a correction lens associated to a specific aberration has an effect on all other aberrations. This cross-coupling complicates precise tuning while also contributing to the observed asymmetry between enhancement settings. Therefore, the provided experimental images should be considered a purely qualitative demonstration of this effect.

\begin{figure}[htp]
    \centering
    \includegraphics[width= 14cm]{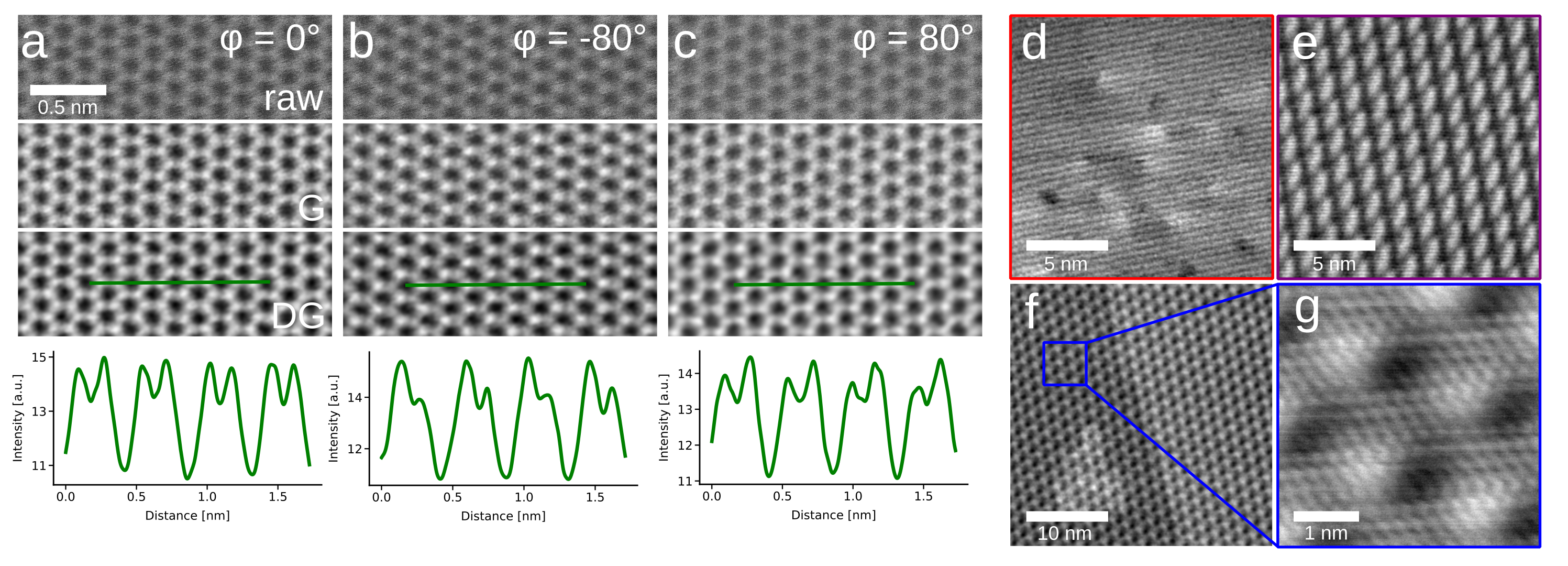}
    \caption{A23-related effects in multilayer hBN. \textbf{a)} Raw, Gaussian-filtered (G) ($\sigma$ = 3 px) and double Gaussian-filtered (DG) (bandpass around first order FFT peaks, inner/outer radius 2.0/10.0~nm$^{-1}$) ADF-STEM images of a 6-layer hBN region without additional A23 aberration as well as a line profile over the double Gaussian-filtered image. The scale bar in the first image is valid for all images. \textbf{b)}  Equivalent images from the same region but with additional  corrector detuning ($R_{23}$ = 132~nm, $\phi_{23}$ = -80\degree). \textbf{c)} Equivalent images from the same region but with different corrector detuning($R_{23}$ = 132~nm, $\phi_{23}$ = +80\degree). \textbf{d), e), f)} ADF-STEM images of the approximately 15~nm thick regions marked by red (f), blue (e) and green (f) stars in Fig.~\ref{Fig 4}a. \textbf{g)} Magnified ADF-STEM image of the interference pattern in panel f.} 
    \label{Fig 5}
\end{figure}

Next, we imaged the six-layer region using the same corrector settings described above, with only the defocus adjusted to optimize image sharpness. Fig.~\ref{Fig 5}a-c show both raw and filtered images of this region in three conditions: without A23 detuning (neutral), with in-phase A23 (enhancing the "B"-site), and with counter-phase A23 (enhancing the "N"-site), using the same amplitude and phase parameters as before. As expected, the baseline setting shows nearly equal contrast between "B" and "N" columns, while the addition of A23 leads to clear contrast enhancement on one side of the atomic column pair, depending on the phase. Here we also observe a similar asymmetry as discussed previously.

Unlike in the monolayer–bilayer step region, where the "neutral" configuration can be reliably found by identifying corrector parameters that remove the contrast difference between neighboring sites in the bilayer, while preserving it in the monolayer, the situation is more ambiguous in thicker regions. Here, the influence of A23 is more insidious, as it can easily mimic stacking-dependent intensity ratios and lead to incorrect conclusions about the number of layers or the orientation of the lattice.

To explore whether A23-related effects persist in significantly thicker regions, we imaged the near-bulk section of the flake, approximately 15~nm thick (ca. 39–48 layers), indicated by the red star in Fig.~\ref{Fig 4}a. Despite using the same correction parameters that produced high-quality images in thinner areas, we were unable to clearly resolve the atomic columns in this region. Manual re-tuning of the corrector settings did not significantly improve image quality, and the resulting image (Fig.~\ref{Fig 5}d) shows poor resolution and diffuse contrast. This is most likely due to a slight sample tilt and could be corrected by tilting the sample in a tilt-holder, which was not available during the experiment. The presence of slight tilt in the thicker regions is also validated by the slight asymmetry of the PACBED features in Supplementary Fig.~\ref{PACBED}.

We repeated this analysis on two additional regions of comparable thickness, whose positions are marked by green and blue stars in Fig.~\ref{Fig 4}a. In both areas (Fig.~\ref{Fig 5}e-f), we observed strong interference patterns typical of tilted samples. Notably, the interference pattern observed in the blue star region (Fig.~\ref{Fig 5}f) resembles that of a monolayer hBN image, albeit with an approximately ten-fold increase in the apparent lattice constant. Fig.~\ref{Fig 5}g provides a magnified view of this pattern, where the hexagonal structure with the actual atomic columns can be clearly discerned within the interference pattern. Interestingly, these interference patterns were remarkably stable and did not change significantly with different corrector settings, albeit changing the defocus led to different pattern geometries in the green star region (Supplemental Materials Fig.~\ref{defocus}). These examples demonstrate that contrast formation in thicker hBN samples is not only dependent on mass-thickness contrast, but also other factors, such as sample tilt, have to be taken into account.

\subsection*{Thickness-dependent cathodoluminescence}
\begin{figure}[htp]
    \centering
    \includegraphics[width= 14cm]{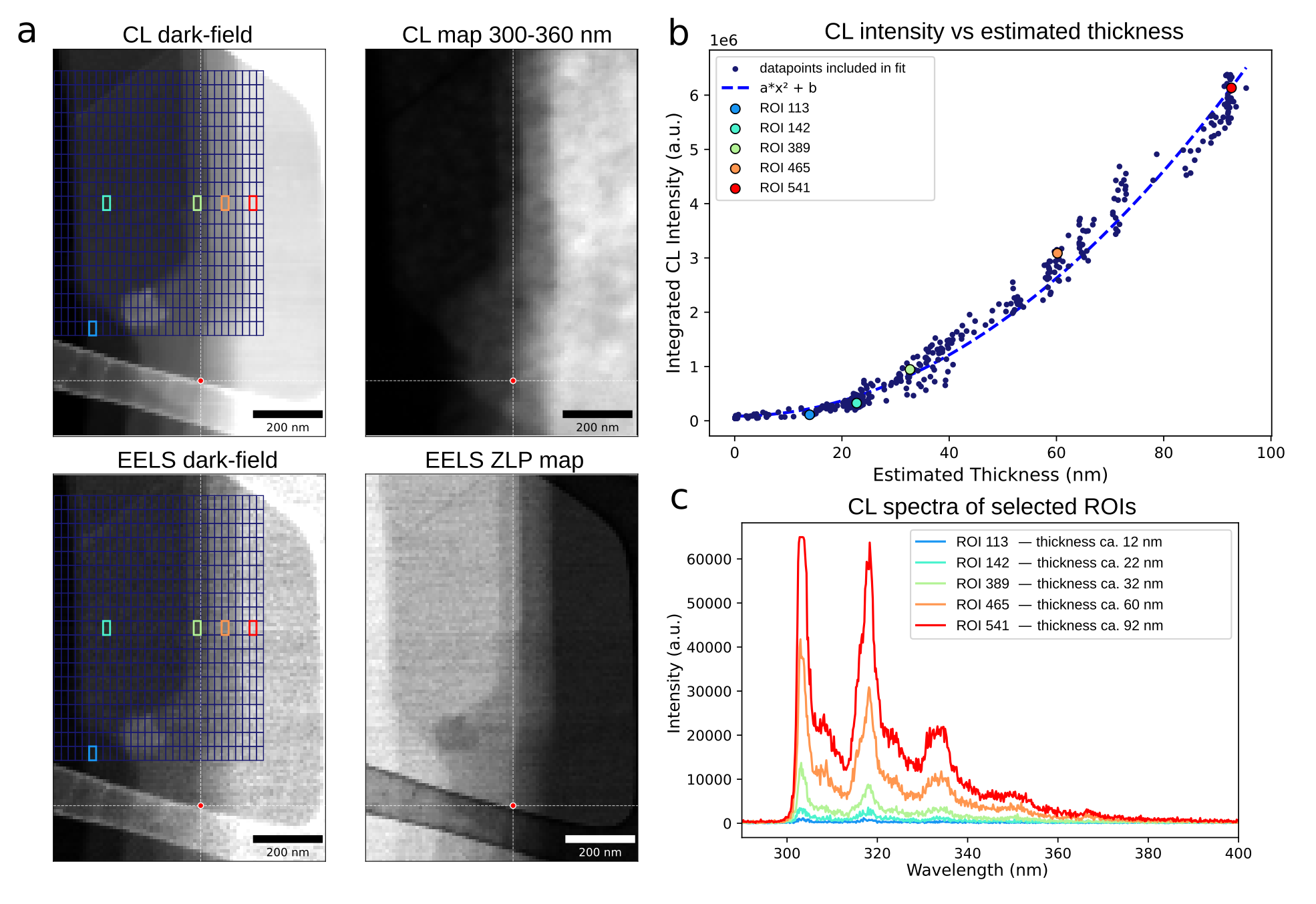}
    \caption{CL emission of multilayer hBN as a function of thickness. \textbf{a)} Top-left: ADF-STEM image corresponding to a CL spectral-image, with the corresponding CL image filtered over a 300 to 360~nm window on the top-right. Bottom-left: ADF-STEM image corresponding to an EELS spectral-image, with the corresponding EELS map filtered over the zero-loss-peak (ZLP) energy region on the bottom-right. CL intensity and estimated thicknesses are estimated for every cell of the overlayed grid. \textbf{b)} CL intensity vs estimated thickness (calculated from the $t/\lambda$ value) for areas shown in panel a, the CL intensity falls off approximately quadratically with thickness. \textbf{c)} Averaged CL spectra for exemplary areas marked with colors in panel a.} 
    \label{Fig 6}
\end{figure}

Having established that defect identification can be complicated due to sample thickness and imaging imperfections, we also show that the cathodoluminescence (CL) emission intensity from defect-engineered hBN samples has a very strong dependence on sample thickness. Cathodoluminescence in STEM is an adequate tool for correlative studies, as it allows for nanometer-scale localization~\cite{tizei_spatially_2013} and measures spectra related to off-resonance photoluminescence \cite{kociak_cathodoluminescence_2017}. For this, crystalline hBN samples with implanted carbon atoms~\cite{taniguchi_synthesis_2007} (see Methods) were sonicated to produce multilayer flakes and drop-casted on lacey carbon TEM grids. On hBN flakes with many terraces (Fig.~\ref{Fig 6}a), CL measurements of the 4.1~eV emission were correlated with EELS measurements which allowed us to estimate the relative thickness of the flake at the probe position by $t/\lambda$ measurement. The CL intensity-over-thickness plot in Fig.~\ref{Fig 6}b shows clearly that the CL intensity drops quadratically with decreasing thickness before reaching the background level defined by readout noise and contributions from transition radiation. Isolated, averaged spectra from specific steps are plotted in Fig.~\ref{Fig 6}c, showing clearly that the emission intensity vanishes at about 10~nm thickness. 

\subsection*{Implications for ADF-STEM imaging}
As we have shown through experiments and image simulations the identification of specific defect types in multilayer hBN using ADF-STEM becomes increasingly unreliable even at modest thicknesses of only a few nanometers. Furthermore, we have thoroughly demonstrated how threefold astigmatism (A23) can artificially alter contrast in ways that mimic true mass-thickness contrast, leading to potentially erroneous interpretations. 

Even in the absence of A23 or other asymmetric aberrations, quantification of ADF-STEM contrast is fundamentally limited by a number of instrumental restrictions and physical effects, such as thermal diffuse scattering, beam coherence, channeling, and detector noise, that introduce blur and variations in intensity. As a result, both the intrinsic "B"/"N" contrast and subtle intensity shifts due to atomic substitutions or vacancies become undetectable at significantly lower layer numbers than predicted by the simulations. While it is difficult to define an universal detectability threshold for a certain feature due to differences in instrumentation, our experience shows that even in monolayer hBN, substitutional carbon atoms are not always reliably detectable, especially when they are located in boron sites and/or located at edges, and when imaging conditions are less than ideal. As a general rule of thumb, it could be considered that an atomic feature becomes indistinguishable from its surrounding once the effective difference in \textit{Z}-contrast to the pristine atomic columns is an order of magnitude lower than the pristine \textit{Z}. 

Another important factor to consider is the effect of image post-processing techniques commonly employed to enhance atomic contrast. In Fig.~\ref{Fig 4} and \ref{Fig 5}, we showed raw experimental ADF-STEM images with and without intentionally introduced A23 aberration. The same images were then processed using a standard Gaussian filter and a double Gaussian filter (often also referred to as inverse-FFT filter). While both methods significantly enhanced the visibility of atomic columns and the apparent "B"/"N" contrast, this enhancement is only reliable if the absence of aberrations such as A23 can be strictly guaranteed. In the presence of such aberrations, contrast enhancement techniques amplify misleading signals, resulting in false-positive identification of an intensity difference between neighboring atomic columns or apparent substitutional atoms. 

A common solution to improve the reliability of substitutional defect identification is the concurrent use of atomic-resolution STEM-EELS~\cite{singla_probing_2024, hou_nanometer_2025}. Indeed, the presence of carbon can be confirmed via its characteristic K-edge, which serves as a strong indicator of local carbon concentration at the beam position. However, while this method is effective in atomically thin samples, its applicability in thicker material is limited because thicker materials lead to larger backgrounds for edges in EELS spectra. Specifically, for carbon in hBN, the tails and fine-structures arising from the plasmon and boron K edge (191~eV) hinders the carbon K edge (284~eV) detection from a single atom. Moreover, in near-bulk hBN, the carbon K-edge signal originating from substitutional atoms is indistinguishable from that of thin and flat airborne hydrocarbon contamination on the sample surface, which itself cannot be identified reliably due to its nearly uniform contrast. Furthermore, the EELS signal could also originate from the residual degradation products of organic polymers on the sample surface, which have been shown to be single photon emitters with spectra similar to the suspected defects~\cite{neumann_organic_2023, toninelli_single_2021, gusdorff_correlated_2025}. Fortunately, such contamination can be effectively removed by annealing the sample at 400°C in ultra-high vacuum (UHV)~\cite{irschik_atomically_2025}.

The significant impact of threefold astigmatism on atomic contrast emphasizes that robust aberration correction is an essential prerequisite for any quantitative image analysis in ADF-STEM. However, even in nominally corrected systems, caution is warranted. Automated aberration correction tools, while effective in many cases, can sometimes give misleading results if too heavily relied upon. Manual adjustments of A23 can also unintentionally amplify artificial contrast shifts, potentially leading to exaggerated "B"/"N" intensity differences, contrast inversion, or even the false appearance of alternative stacking orders. However, in some cases a certain degree of aberration may even be advantageous for the analysis, for example when the contrast enhancement permits the discernment of a lower \textit{Z} element from its higher \textit{Z} neighbors. Importantly, the same caveats apply to other layered materials with hexagonal symmetry. In 2H-phase transition-metal dichalcogenides, residual A23 can exaggerate the contrast between transition metal and chalcogen sites, causing the images to resemble those of a 1T phase~\cite{lopatin_aberration-corrected_2020}. In graphene, even minor threefold astigmatism can make it appear visually similar to monolayer hBN~\cite{sawada_accurate_2017, hofer_atom-by-atom_2021}. 

While we here focus on the effect of A23, it is important to acknowledge that, in reality, there is always an interplay between multiple probe aberrations. The contrast-transfer function is highly non-linear, and visually correcting one type of aberration can simultaneously amplify the impact of others. Most importantly, even if an image appears “correct” and exhibits the expected atomic resolution, this does not guarantee the absence of residual aberrations, as a certain amount of aberration is still consistent with typical criteria for "good" resolution (e.g., the D50 and $\pi/4$ criteria)~\cite{haider_upper_2000, sagawa_exploiting_2022}. 

It is also essential to compare the intensity ratios obtained under different instrumental conditions. In this study, the simulation parameters were selected to allow direct comparison with experimental data acquired using the 60 kV Nion UltraSTEM 100, which operates with a convergence semi-angle ($\alpha$) of 35 mrad, a medium-angle annular dark-field (MAADF) collection range of 60–200~mrad, and a high-angle annular dark-field (HAADF) range of 80–300~mrad. These settings are atypical, as most aberration-corrected instruments employ smaller convergence angles, typically around 20-30~mrad, with inner MAADF detection angles near 40–60~mrad and inner HAADF detection angles above 90~mrad. A common rule of thumb defines the MAADF regime as approximately 2–3$\times \alpha$ and places the start of the HAADF regime to at least 3$\times \alpha$~\cite{hartel_conditions_1996}. Therefore, the HAADF detector configuration used in our setup is rather unconventional, encompassing much of the angular range usually associated with a typical MAADF detector. Supplementary material Fig.~\ref{Fig: comparison} illustrates how the intensity ratio evolves both in the aberration-free case and when a 50~nm in-phase A23 aberration is introduced, using instrumental parameters comparable to those employed in other single-photon emitter studies. These simulations clearly demonstrate that the thickness-dependent intensity ratios behave similarly in all cases, but it can be observed that A23 has a significantly larger effect in the MAADF regime, where phase-contrast has a stronger contribution. It should be emphasized, that the provided image simulations cannot perfectly capture the signal-to-noise characteristics at higher scattering angles, since dose limitations are introduced only by adding Poisson-distributed noise to the simulated images. Consequently, the reduced scattering intensity at higher angles and the resulting practical limitations of HAADF detectors in real instruments are not fully represented in the simulated data.

\subsection*{Implications on quantum emitter studies}
The impact of the presented limitations of ADF-STEM imaging is particularly significant for studies aiming to identify the atomic origin of single-photon emission in hBN. While computational investigations have convincingly linked various point defects, such as vacancies, interstitials, and substitutional dopants, to potential single photon emission, experimentally matching a specific emission line to a distinct atomic defect remains far more complicated than is often assumed.

In typical photoluminescence studies, the spatial resolution of the optical system limits the excitation and collection to an area of at least 100~nm$^2$, encompassing thousands of individual atoms in dozens of layers, and potentially a large variety of defect types. These defects may exist both near the surface and deeper within the bulk, making it nearly impossible to attribute the emission to a single, isolated structure. Reducing the material to a monolayer or a thickness of only a few layers, and combining STEM with optical near field techniques could improve this correlation. However, care must be taken not to alter the hBN lattice, which can be easily damaged during STEM imaging~\cite{bui_creation_2023}. The more suitable alternative for very thin hBN samples is measuring the emission in scanning tunneling luminescence  microscopy (STML)~\cite{huberich_atomically-resolved_2025}, or to combine STM with optical near field techniques~\cite{roelcke_ultrafast_2024}.

Given the limitations of HAADF-STEM analysis in thicker samples, it is highly desirable to obtain strongly localized emission spectra from significantly thinner specimens. As demonstrated in Fig.~\ref{Fig 6}, discernible CL spectra can still be recorded from hBN flakes about 10~nm thick, and several studies have reported photoemission from even thinner samples~\cite{liu_single_2025, fernandes_room-temperature_2022}. In the most extreme case, Tran et al. observed vacancy-related single-photon emission from a single layer of hBN~\cite{tran_quantum_2016}. The reduced emission intensity in thinner samples can be attributed to several factors, including a smaller number of potential emitters~\cite{durand_optically_2023}, a reduced excitation volume~\cite{clua-provost_impact_2024}, weaker dielectric screening~\cite{li_dielectric_2015}, substrate-induced quenching~\cite{qiu_atomic_2024}, and faster photobleaching~\cite{li_prolonged_2023}. Nonetheless, there are strategies to enhance emission from atomically thin samples. These include cooling the sample below 100 K~\cite{akbari_temperature-dependent_2021} or exploiting exciton trapping in twisted hBN heterostructures~\cite{roux_exciton_2025}. Moreover, advanced photoemission spectroscopy techniques may enable the localization of emission sites down to the atomic scale~\cite{duan_bayesian_2025}.

\section*{Conclusion}
Our results demonstrate that atomic-resolution identification of defects in multilayer hBN using ADF-STEM is inherently limited by material thickness and residual aberrations. Even under near-ideal conditions, boron–nitrogen contrast becomes indistinguishable beyond approximately 17 layers, while substitutional carbon atoms rapidly lose visibility at far smaller thicknesses. Residual non-radially symmetric aberrations, particularly threefold astigmatism, can further introduce artificial contrast that mimics genuine structural features, leading to potential misinterpretation of defect types and stacking sequences. These effects become especially problematic when studying near-bulk hBN flakes commonly employed in photonic investigations.

Our findings emphasize the need for caution when correlating single-photon emission with specific atomic-scale defects in hBN. Reliable structural attribution requires atomically thin samples and complementary techniques, such as STEM-EELS or scanning probe microscopy, used under carefully controlled conditions. Ultimately, recognizing the intrinsic imaging limits of ADF-STEM is essential for accurate defect identification and for advancing a consistent microscopic understanding of quantum emitters, not only in hBN, but in all layered materials.

\section*{Acknowledgments}
We would like to thank Kenji Watanabe and Takashi Taniguchi for providing the hBN samples used in the cathodoluminescence experiments. This project has been funded in part by the National Agency for Research under the program of future investment TEMPOS-CHROMATEM (reference no. ANR-10-EQPX-50), the Vienna Doctoral School in Physics and the Austrian Science Fund (FWF) [10.55776/P35318, 10.55776/DOC142, 10.55776/COE5 and 10.55776/P34797]. For open-access purposes, the author has applied a CC-BY public copyright license to any author-accepted manuscript version arising from this submission.

\section*{Authors Contributions}
DLam and SC conceived the study. DLam, SC, MM, JD and BM conducted the experimental work and performed the data analysis at the University of Vienna with JK. AMF and LHGT conducted the experimental work and performed the data analysis at the Université Paris-Saclay, which were further discussed with MK. SC and DLor prepared the hBN samples. DLam, SC, BM, AMF, MM, LHGT, JK, and LF co-wrote the manuscript with contributions from all authors. LF, LHGT and JK supervised the study.

\section*{Competing interests}
The authors declare no competing interests.

\section*{Data and material availability}
All data underlying this study will be made available through University of Vienna PHAIDRA repository upon acceptance of the manuscript.

\bibliographystyle{elsarticle-num} 

\bibliography{references}

\renewcommand{\appendixname}{Supplementary Material}
\renewcommand{\thefigure}{S\arabic{figure}} \setcounter{figure}{0}
\renewcommand{\thetable}{S\arabic{table}} \setcounter{table}{0}
\renewcommand{\theequation}{S\arabic{table}} \setcounter{equation}{0}

\section*{Methods}
\subsection*{Image simulations}
ADF-STEM image simulations were performed based on independent atom models of pristine and defective hBN using the \textit{ab}TEM package~\cite{madsen_abtem_2021}. In all cases parameters were set to replicate common imaging conditions of Nion probe-corrected microscopes with an electron beam energy of 60 keV, a probe convergence angle of 35 mrad, an ADF semi-angular range of 80-300~mrad, a source size of 0.03~nm and a dose of $10^6$~$e^-$/\AA$^2$. To account for thermal diffuse scattering, we implemented the frozen-phonon model with at least 10 snapshots per image using a standard deviation of atomic displacements of 0.1~\AA. All simulations had a cell size of at least 8$\times$5 unit cells, a grid sampling of 0.3~\AA, a slice thickness of 0.5~\AA~and a potential sampling of 0.05~\AA. The intensity ratios were derived from the maximum value at every atomic position, all atomic columns within 0.4~nm from the border of the simulation have been excluded, "B" and "N" columns had both at least 25 datapoints per image, $\mathrm{C_B}$ and $\mathrm{V_B}$ columns had 7 datapoints per image. The confidence intervals have been calculated based on the lowest number of datapoints for a column type and on the standard deviation of the ratios.

\subsection*{Sample preparation}

\textbf{Mechanical Exfoliation}
The multilayer sample was created using micro-mechanical exfoliation of bulk hBN. A flake of hBN was carefully pushed against sticky exfoliation tape so that the top few layers are attached. The tape was then folded and peeled multiple times (ca. 12 times) to exfoliate the individual layers. The flakes were then transferred onto a SiO$_{2}$ chip via thermal adhesion on a hot plate. A flake of desired thickness was selected by its optical contrast in a visible light microscope and a Quantifoil (QF) gold (Au) TEM grid was then thermally adhered onto the SiO$_{2}$ region with isopropyl alcohol (IPA) and heating. The SiO$_{2}$ chip was detached via chemical etching using a couple of drops of concentrated KOH solution, resulting in the flake being transferred onto the TEM grid.\\
\textbf{Electrochemical delamination}
Using CVD-grown monolayer hBN on copper (Cu) foil (purchased from Sigma-Aldrich), the hBN sample was prepared using the electrochemical delamination method~\cite{sun_mechanism_2015,chen_wafer-scale_2020}. The foil was cut into a small square piece that was roughly the size of a TEM grid, and was spin-coated with 2 drops of polymethyl methacrylate (PMMA). After curing on a hot plate at 150$^{\circ}$C for ca. 30 min, less than 1 mm was cut off from the edges of the piece to remove PMMA spillover. Using a 1 M NaOH solution as the electrolyte, a platinum (Pt) wire as the anode and the Cu foil (PMMA/hBN/Cu stack) as the cathode, the production of H$_{2}$ gas at the copper gently delaminates the PMMA/hBN stack using 4~V from a power supply. The delaminated stack was washed in a deionized (DI) water bath to remove electrolyte residue and was then fished onto a custom Ti coated SiN TEM grid with a membrane hole size of 1 $\micro$m. The sample was then heated to 150$^{\circ}$C on a hot plate for ca. 30 min to evaporate any trapped liquids and promote adhesion. The PMMA was removed by immersing the sample into a hot acetone bath (50$^{\circ}$C) for ca. 1 h, followed by a room temperature isopropyl alcohol (IPA) bath for ca. 30 min to complete the sample preparation process.
\\
\subsection*{Microscopy at the University of Vienna}
All microscopy data in Figs.~\ref{Fig 2},~\ref{Fig 4} and \ref{Fig 5} have been acquired with a Nion UltraSTEM 100 microscope located at the University of Vienna with an electron beam energy of 60 keV and a beam current of ca. 140 pA. Images have been recorded using the MAADF detector with an angular range of 60-200 mrad. The microscope is equipped with a C3/C5 aberration correction system and the automated correction tool "AutoSTEMX". After automated correction, aberrations up to third order have been fine-tuned manually. Artificial aberrations have been introduced by detuning the aberration correction in the corrector interface.

\subsection*{Carbon implantation and EELS measurement}

For carbon implantation, approximately 1.5$\times$ $10^{-8}$ Torr of methane (CH$_4$) was introduced into the Nion microscope column (base pressure ca. 1$\times$ $10^{-10}$ Torr) through a leak valve. The number of CH$_4$ molecules impinging on the sample surface can be estimated using

\begin{equation} \label{arealimp}
J = \frac{2p}{\sqrt{2\pi m k_B T}},
\end{equation}
where $p$ is the gas pressure, $m$ is the molecular mass, $k_B$ is the Boltzmann constant, and $T$ is the temperature. Using Eq. \eqref{arealimp}, the areal impingement rate of methane molecules ($m$ = 2.7 $\times$ 10$^{-26}$ kg) at T = 300K and a pressure of 1.5 $\times$ 10$^{-7}$ Torr is calculated to be approximately 0.76 s$^{-1}$nm$^{-2}$. We point out, that the actual pressure at the sample may be as much as one order of magnitude higher than the value reported by the vacuum gauge.

Substitutional carbon atoms have been introduced into the multilayer hBN sample by scanning the field of view repeatedly over a time of about 2 min. After the first changes in atomic composition became apparent, EELS maps with 64$\times$64 pixels and a pixel dwell time of 50 ms have been acquired using a Gatan PEELS 666 spectrometer with an Andor iXon 897 CCD camera and an energy-dispersion of 0.6 eV/pixel. In all cases, the EELS spectra have been acquired in the range between ca. 170 and 440 eV electron loss, encompassing the full K-edges of boron, carbon and nitrogen. The background of all spectra has been subtracted using a power-law fit. After acquisition of EELS spectra, a MAADF image with 512$\times$512 pixels has been acquired at the approximate start position of the EELS map.

\subsection*{CL measurements and thickness estimation}

CL maps and corresponding EELS maps of multilayer hBN flakes exfoliated from carbon-doped bulk crystals synthesized by the high-temperature and high-pressure method~\cite{taniguchi_synthesis_2007} were acquired at the LPS Orsay with a Nion Hermes microscope equipped with an Attolight parabolic mirror for CL light collection. The used acceleration voltage was 60 kV, during the measurements the sample was cooled with liquid nitrogen to approximately 100 K. The CL signal was coupled into a fiber bundle and led to a Princeton optical spectrometer where it was dispersed by a grating and the signal was detected on a ProEM EMCCD camera with 1600x200 pixels. The EELS signal was acquired using a Nion spectrometer and an event-based direct electron detector consisting of 4 Timepix3 chips (CheeTah from ASI).
CL and EELS maps have been acquired on two different areas with multiple steps of different thicknesses on two different hBN flakes. For the first area, the CL map was acquired with 10~nm pixel size and 100 ms dwell time, the EELS  with 10~nm pixel size and 1 ms dwell time. For the second area, the CL map was acquired with 5~nm pixel size and 100 ms dwell time, the EELS  with 2.5~nm pixel size and 1 ms dwell time.
Data treatment was performed in the following manner: For CL, first scale and offset of the energy axis were fixed by comparing a spectrum of a reference lamp taken with its known peak values. For EELS the offset of the energy was fixed by setting the zero to the position of the zero loss peak. The spatial coordinate systems of the respective CL and EELS maps were aligned by identifying features that are well discernible in the the dark field images taken during the acquisition of the CL and EELS maps and adapting offset and scale so that those features have the same coordinates in both maps. With a common coordinate system established, the overlapping portions of the CL and EELS maps were sliced into a grid with small patches. For each patch the averaged CL intensity was integrated in the range of 300~nm to 360~nm, corresponding to the range of the 4.1 eV emission, and from the averaged EELS spectrum the sample thickness was roughly estimated via the $t/\lambda$ method~\cite{malis_eels_1988} using a value of 50~nm for the inelastic mean free path of 60 keV electrons in hBN calculated from equations 5.2 and 5.2a and using data from table 5.2 provided in~\cite{egerton_electron_2011}. Then the vacuum intensity, which is mainly associated with readout noise, was subtracted from the CL intensity. The remaining background is mainly due to transition radiation, which is emitted whenever a fast charge crosses a dielectric boundary~\cite{malis_eels_1988} and therefore is difficult to avoid in CL. The $t/\lambda$ method provides merely a rough estimation of the local sample thickness with a precision of $\pm$ 20 \% at best, additionally any contamination present on the surface of the flake will contribute to the measured thickness and leads to further imprecision~\cite{malis_eels_1988}. 

\subsection*{Atomic scale EELS data}
\begin{figure}[htp]
    \centering
    \includegraphics[width= 12cm]{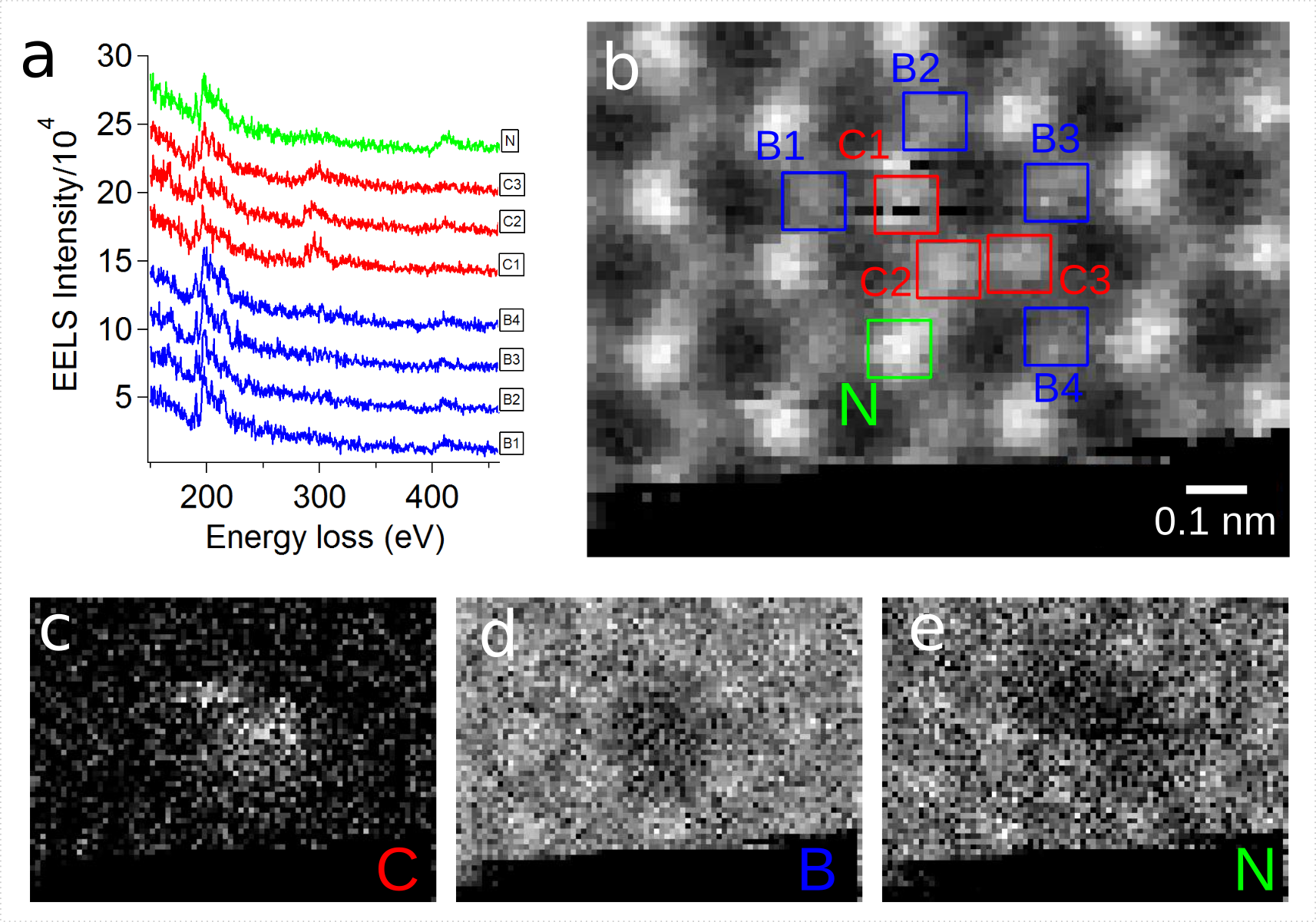}
    \caption{\textbf{a} EELS spectra averaged over the rectangular regions in panel b. \textbf{b} Corresponding HAADF-STEM image of monolayer hBN. The rectangular regions indicate atomic position corresponding to the EELS map intensities. \textbf{c} EELS map integrated over the carbon K-edge (180-240 eV). \textbf{d} EELS map integrated over the boron K-edge (280-340 eV). \textbf{e} EELS map integrated over the nitrogen K-edge (390-510 eV). The small, three-carbon atom inclusion, was not present in the original sample. It formed when a three-atom-wide void was created and subsequently filled by three carbon atoms, similarly to what has been described previously by Liu et al.~\cite{liu_postsynthesis_2016}}. 
    \label{Luiz EELS}
\end{figure}

The ADF/EELS data in Fig.~\ref{Luiz EELS} was acquired on the Nion UltraSTEM 200 located at the LPS-Orsay operated at 60 keV. The beam convergence half angle was 34 mrad and the EELS collection half angle was 60 mrad (2 mm spetrometer entrance aperture). The EELS spectra were acquired on a modified Gatan spectrometer equipped with a thin scintillator lens-coupled to a back-illuminated electron enhanced charge coupled device (ProEM EMCCD from Princeton Instruments) with 1600$\times$200 pixels. The sample for these experiments were hBN monolayer grown by CVD on copper. 

\subsection*{PACBED and EELS thickness estimation}
4D-STEM data was recorded with the Nion UltraSTEM 100 at the University of Vienna using a convergence angle of 15.92 mrad and an acceleration voltage of 60 kV. The Dectris ARINA detector (192$\times$192 pixels) was configured with a maximum collection angle of 109 mrad. Data collection was performed over 256$\times$256 real-space scan positions with an FOV of 2$\times$2~nm and a dwell time of 50 \textmu s. PACBED patterns were obtained by subsequently averaging in real-space. PACBED patterns of known thickness and tilt were produced by simulating 4D-STEM data with \textit{ab}TEM using the microscope parameters mentioned above, a cell size of 4~nm side length, a grid and potential sampling of 0.05 ~\AA~ and a slice thickness of 1 ~\AA~. Computations were performed increasing layer-wise up to 24 layers and then in steps of 2 layers to 66 layers of hBN, always including a tilt series of 0, 50 and 100 mrad. The processing steps consist of data interpolation to match the grid size of the detector and averaging the diffraction patterns over all real-space positions. To match PACBED patterns \cite{lebeau_position_2010}, all images were visually aligned and normalized by subtracting the mean and dividing by the standard deviation. A mask was applied to the zero-order PACBED spot to make the similarity assessment more sensitive to the features of higher-order PACBED spots. The structural similarity index measure (SSIM) as implemented in the scikit-image library, normalized cross correlation (NCC) and the mean square error (MSE) were utilized to determine the similarity of experimental and simulated PACBED patterns. The best match is based only on the SSIM.

EELS measurements were performed with a Gatan PEELS 666 spectrometer, an Andor iXon 897 CCD camera using an energy dispersion of 0.16 eV/px, a dwell time of 0.5 ms, and the accumulation of 10000 spectra for each position. Data collection was carried out multiple times within each region of interest. The $t/\lambda$ method \cite{iakoubovskii_thickness_2008} was implemented using the hyperspy package~\cite{pena_hyperspyhyperspy_2024}. The effective mean-free-path (MFP) was estimated by fitting the data collected on the 2 layer and 6 layer region. In addition, a rough estimate of the inelastic mean-free-path was calculated using the equations in Ref. \cite{egerton_electron_2011}. The thickness was estimated for all regions using these results. Estimates were averaged for each region.

\section*{Supplementary Figures:}

\begin{figure}[htp]
    \centering
    \includegraphics[width= 14cm]{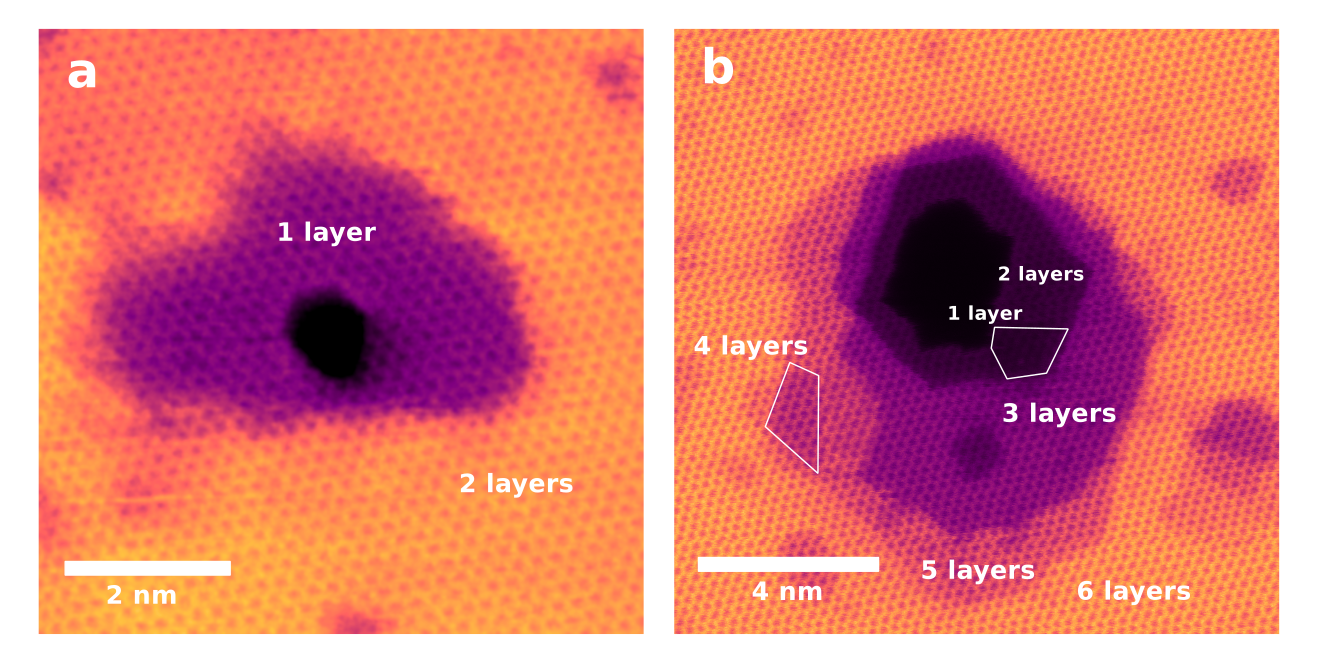}
    \caption{\textbf{a} MAADF-STEM image of the bilayer hBN region with a hole drilled to show the number of layers. \textbf{b} The same for a six-layer region.} 
    \label{drilling}
\end{figure}

\begin{figure}[htp]
    \centering
    \includegraphics[width= 14cm]{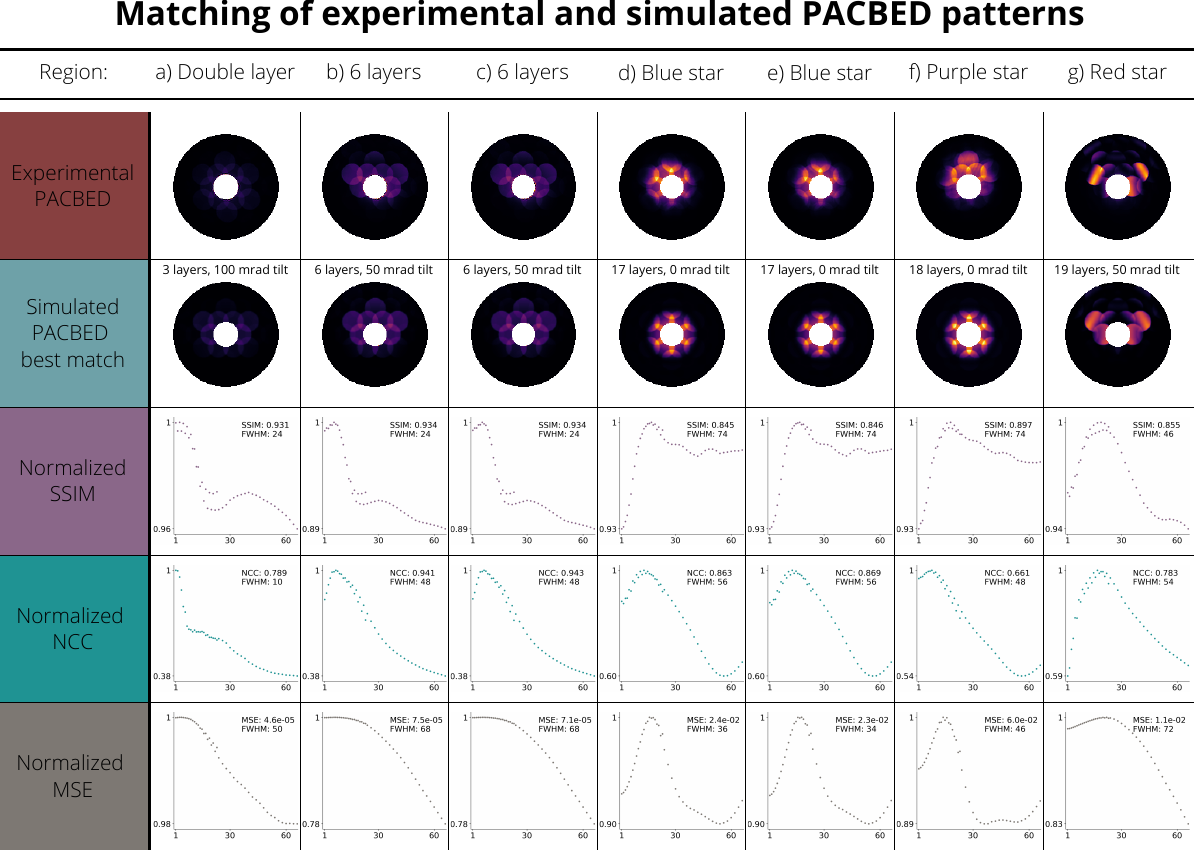}
    \caption{\textbf{a-g} Comparison of experimental and best fitting simulated PACBED patterns as well as three normalized image similarity criteria, similarity index measure (SSIM), normalized cross correlation (NCC) and mean square error (MSE), as a function of thickness for different areas on the sample indicated in the optical microscopy image in~\ref{Fig 4}a.} 
    \label{PACBED}
\end{figure}

\begin{figure}[htp]
    \centering
    \includegraphics[width= 14cm]{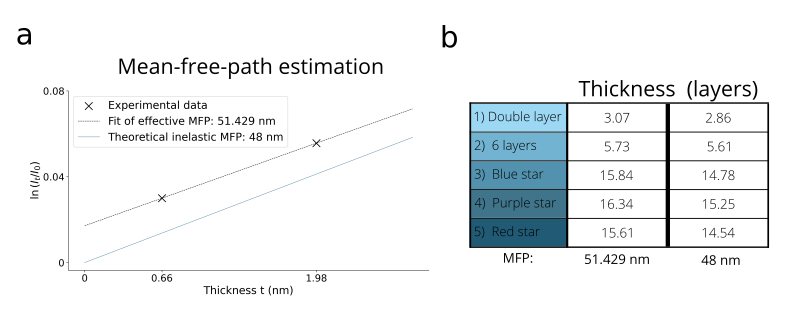}
    \caption{Thickness estimation using the $t/\lambda$ method \textbf{a} Calibration of the $t/\lambda$ curve using the known 2- and 6-layer regions \textbf{b} Table with estimated thicknesses for the regions indicated in the optical microscopy image in~\ref{Fig 4}a using both the theoretical and fitted mean free path.} 
    \label{EELSthickness}
\end{figure}

\begin{figure}[htp]
    \centering
    \includegraphics[width= 14cm]{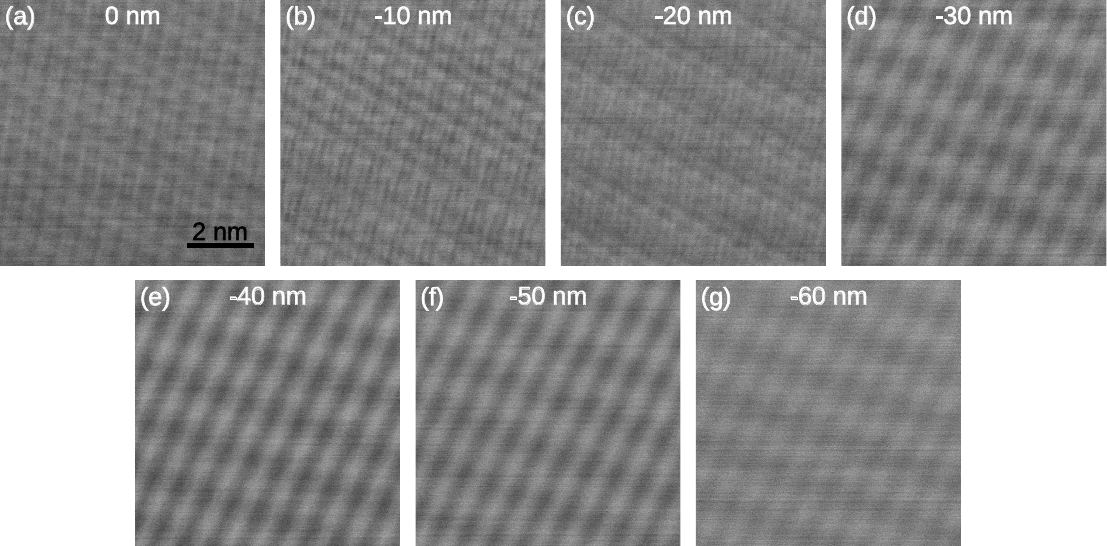}
    \caption{MAADF-STEM images of the multilayer region indicated with a blue cross in Fig.~\ref{Fig 1}a at different defocus values.} 
    \label{defocus}
\end{figure}

\begin{figure}[htp]
    \centering
    \includegraphics[width= 14cm]{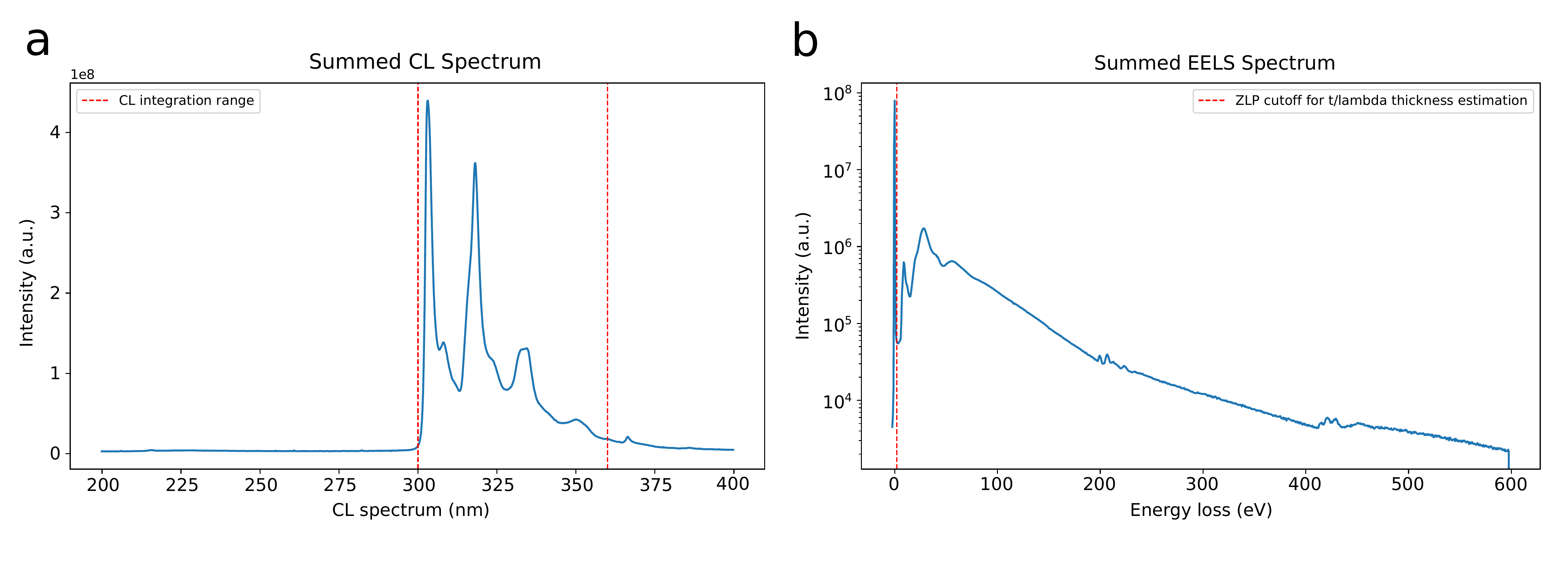}
    \caption{\textbf{a} CL spectrum summed full area with integration limits for calculation of integrated CL intensity shown as red vertical lines. \textbf{b} EELS spectrum (logarithmic intensity scale) summed over the full area with the maximum energy of the  ZLP area marked by a red line.} 
    \label{Cl spectrum fig 6}
\end{figure}

\begin{figure}[htp]
    \centering
    \includegraphics[width= 14cm]{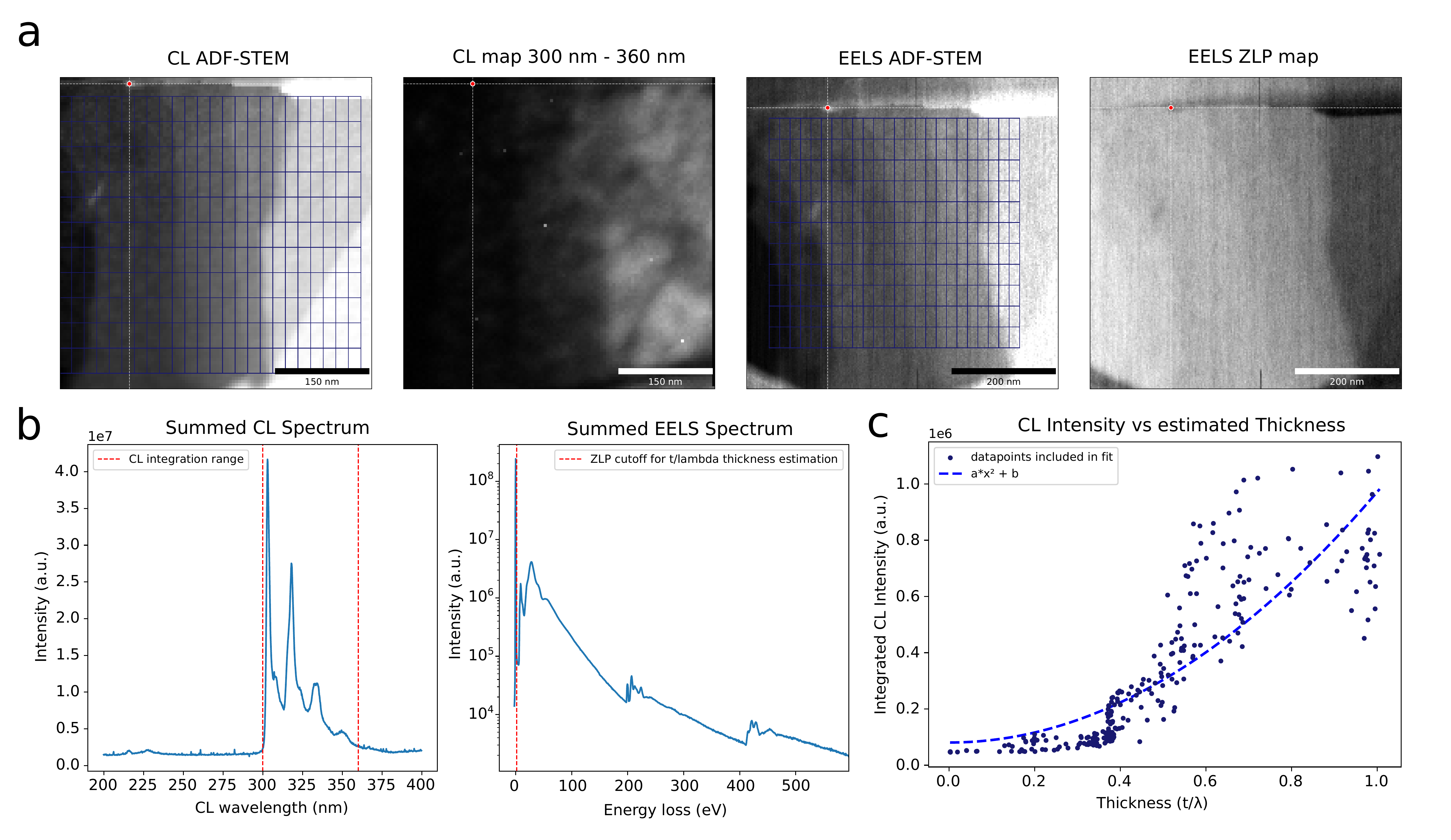}
    \caption{Thickness dependent CL of another flake \textbf{a} CL map and corresponding DF image and EELS map and corresponding ADF-STEM image (ADF-STEM image contrast adjusted between 0th and 95th percentile) of same area on hBN flake, with same coordinate system shown in white with coordinate origin in red. \textbf{b} Summed CL spectrum of the full area with integration limits for integrated CL intensity calculation shown as red vertical lines. EELS spectrum (logarithmic intensity scale) summed over full area with delimitation of ZLP area and energy loss area as used in t/$\lambda$ method for thickness estimation shown as red vertical lines. \textbf{c} CL intensity vs estimated thickness (t/ $\lambda$) for the area shown in a, the CL intensity falls off approximately quadratically with thickness.} 
    \label{Fig: another flake}
\end{figure}

\begin{figure}[htp]
    \centering
    \includegraphics[width= 14cm]{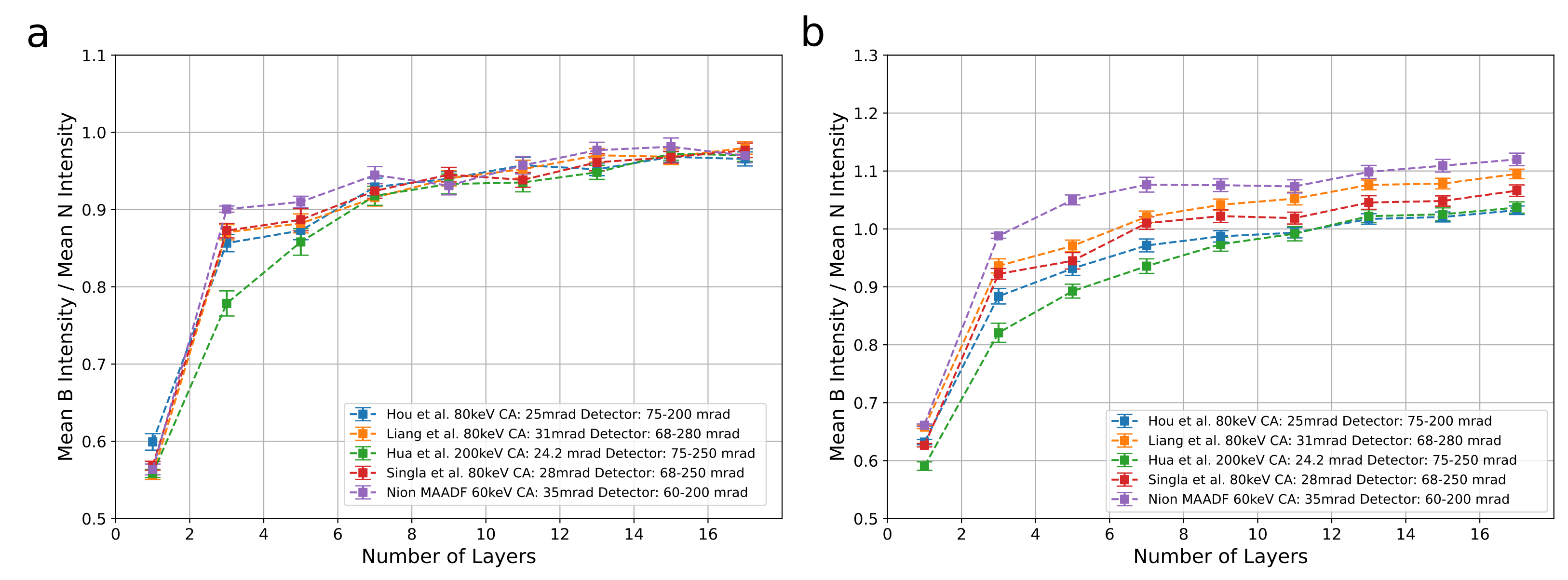}
    \caption{\textbf{a} B/N intensity ratios as a function of thickness for different experimental setups in correlative studies. For Hou et al.~\cite{hou_nanometer_2025} only the type of the microscope (Thermo-Fisher Spectra 300), detector (HAADF) and the beam energy (80 keV) were given. Convergence angle (25 mrad) and detector angles (75-250 mrad) were assumed based on typical values for this microscope used in studies of the same group (e.g. Ref.~\cite{ji_nanocavity-mediated_2024}). Hua et al. used a similar setup (Thermo-Fisher Spectra 200) as Hou et al. but with 200 keV beam energy, all other parameters had to be assumed. For Liang et al.~\cite{liang_site-selective_2025} and Singla et al.~\cite{singla_direct_2025, singla_probing_2024} all values could be found in the respective publications. Nion MAADF parameters reflect the MAADF setting used at the Nion UltraSTEM 100 in Vienna. \textbf{b} Equivalent graph with 50~nm A23, in-phase with the crystal orientation.} 
    \label{Fig: comparison}
\end{figure}



\end{document}